\documentclass[12pt]{article}
\usepackage{epsfig}
\usepackage{a4wide}

\newcommand{\lab}[1]{\label{#1}}
\newcommand{\cit}[1]{\cite{#1}}

\newcommand{\beq}{\begin{equation}}
\newcommand{\eeq}{\end{equation}}
\newcommand{\bea}{\begin{eqnarray}}
\newcommand{\eea}{\end{eqnarray}}

\newcommand{\e}{\mbox{e}}
\newcommand{\s}{\sigma}
\newcommand{\be}{\beta}
\newcommand{\ta}{\tau}
\newcommand{\rh}{\rho}
\newcommand{\n}{\nu}
\newcommand{\om}{\omega}
\newcommand{\la}{\lambda}
\newcommand{\G}{\Gamma}

\renewcommand{\a}{\alpha}

\newcommand{\del}{\delta}

\newcommand{\hp}{\hat p}
\newcommand{\hx}{\hat x}
\newcommand{\hD}{\hat D}

\newcommand{\bk}{{\bf k}}
\newcommand{\bx}{{\bf x}}
\newcommand{\by}{{\bf y}}
\newcommand{\bp}{{\bf p}}

\newcommand{\bP}{{\breve \Phi}}

\newcommand{\ct}{{\cal T}}

\newcommand{\cW}{{\cal W}}

\newcommand{\ccl}{{ l}}

\newcommand{\nn}{\nonumber}
\newcommand{\ad}{a^\dagger }
\newcommand{\as}{\alpha^*}
\newcommand{\1}{\mbox{1 \hspace*{ -1em} 1  }}
\newcommand{\N}{\mbox{ N\hspace*{ -1.25em} I \hspace*{ 0.1em} }}
\newcommand{\9}{\partial}
\newcommand{\de}{\delta}
\newcommand{\dd}{\raisebox{0.03em}{:} }
\newcommand{\lb}{\lbrace}
\newcommand{\rb}{\rbrace}

\newcommand{\tr}{\mbox{tr}}

\begin{document}
\rightline{TUW-96-29}
\rightline{hep-th/9703105}
\rightline{\today }

\bigskip \bigskip \bigskip \bigskip

\begin{center} 
{\Huge Wigner Functionals and their Dynamics \\ 
in Quantum-Field-Theory}

\bigskip \bigskip \bigskip \bigskip

Herbert Nachbagauer\footnote{email: herby \@@ tph16.tuwien.ac.at}\\

\bigskip
 
Technische Universit\"at Wien \\
Wiedner Hauptstra{\ss}e 8-10 \\
A-1040 Wien, Austria \\
\end{center}

\bigskip \bigskip

\abstract{We reformulate time evolution of systems in mixed 
states in terms of the classical observables of correlators 
using the Weyl correspondence rule. 
The resulting equation of motion for the Wigner functional of 
the density matrix is found to be of 
the Liouville type. To illustrate the methods developed,
we explicitly consider a scalar theory with quartic self-interaction 
and derive the short time behaviour with the non-interacting thermal 
density matrix as initial condition. In the scalar case, the complete 
correlator hierarchy is studied and restrictions are derived 
for spatially homogeneous initial conditions and systems with unbroken 
symmetry. 
\\
\smallskip

 }

\newpage

\section{Introduction}
The study of statistical ensembles of quantized fields out of thermal 
equilibrium has a number of interesting applications. Prominent
examples are  
the study of hot QCD matter in heavy ion collisions or the behaviour 
of matter in the course of cosmological phase transitions. 
In contrast to equilibrium theory,
where considerable success has been achieved in recent years \cit{equil},
the theoretical framework for the description of processes far from 
equilibrium still lacks a systematic basis. Although some attempts 
have been made to derive the first few stages of a cut BBGKY-hierarchy
on the basis of equal time correlators \cite{thom} and using other 
methods \cite{wett}, a comprehensive understanding of the interrelation
between observables and the abstract Schr\"odinger evolution for 
systems of quantized fields is still incomplete. 

Physics of non-equilibrium processes typically involves tree elements.
Firstly, the description of an initial state by means of a number of
-- in principle measurable -- observables. Secondly, time evolution 
defined by a dynamical equation together with the corresponding 
Hamiltonian operator, and third, the observation of some 
quantities at later times. The second and third step can be discussed 
on the basis of fundamental principles. In particular, time evolution 
of quantum mechanical systems is encoded in the von Neumann equation which 
determines the time evolution of the density matrix operator.     
Integrating time evolution to later times, it is straight-forward to 
extract expectation values for operators. Only the first step 
requires a statistical assumption about the system, since complete 
knowledge of the initial density matrix is usually not accessible.  

In this paper we focus on the dynamical aspects of statistical 
ensembles of quantized fields but reformulate time evolution in terms 
of classical quantities instead of operator evolution. This casts the 
theory in a form which appears more natural when one is interested in 
observables. Using the effective action as dynamical object generating 
equal time correlators, this problem has been discussed recently 
\cite{nach}.

Based on the Weyl correspondence principle, we consider the classical 
counterparts for quantum mechanical operators and the density matrix. 
Starting with the von~Neumann equation, we focus on the construction of 
a dynamical equation  of the Liouville type for the classical counterpart 
of the density matrix operator. In order to illustrate the abstract result, 
we derive an expression for the Weyl transform 
of the non-interacting equilibrium density matrix and use it as an 
initial condition for the time evolution in an interacting system. 
We find that mass resummation is necessary in order to obtain a 
sensible physical result. Furthermore, we explicitly derive the hierarchy 
of equations of motion for general initial conditions for the model 
of a scalar field with quartic interaction. Particular conditions on 
field correlators for systems which exhibit spatial homogeneity and 
with unbroken symmetry are studied systematically. 

\section{Quantum-mechanics}

\subsection{General}
In order to outline the concepts applied, we demonstrate the basic
steps using a zero-dimensional quantum mechanical system as example. 
A harmonic one-dimensional oscillator is defined by the Hamiltonian 
$  H = \omega a^{\dag }a $ 
which Fock-space is spanned by the eigenstates of $H$.
Systems in mixed states are described by a hermitian density matrix 
operator $ \hat \rho $, normalized to unity. 
Expectation values of operators $ \hat A $ are given by the trace
$ <\hat A> = \mbox{tr} \hat \rho \hat  A $
which is independent of the basis chosen. It can be calculated in 
energy-eigenstates or in position eigenstates as well, 
\beq <\hat A> = \sum _n   < n | \hat \rho \hat A |n >=
\int dx < x |  \hat \rho \hat A  | x >.
\eeq 

\subsection{Weyl-correspondence principle}

Each operator is sub-ordinated a classical observable in a state 
by taking the trace of the product with the density operator. 
An observable defined in such a way 
depends on the particular mixed state of the system under consideration.
One may thus ask if it is possible to define the corresponding 
classical quantity independent of any particular density matrix, 
which in a more universal way permits to calculate observables in a 
given state. This is in fact possible by the Weyl correspondence 
rule \cit{weyl}. Suppose we can decompose an operator as 
\beq 
\hat A  = \int d\sigma d \tau \tilde A ( \sigma, \tau )
\e^{i \sigma \hat p + i \tau \hat x } ,
\eeq
where $\hat x $ and $\hat p $ are the canonical position and momentum 
operators resp., the corresponding classical function is defined to 
have the same Fourier transform, 
\beq 
A (p,x) = \int d \sigma d \tau  \tilde A  ( \sigma, \tau ) 
\e^{i \sigma p + i \tau  x } \lab{toclass}.
\eeq
So far, this is only a definition of the classical counterpart.
In order to justify it, we consider the quantum-mechanical expectation 
value of an operator in a mixed state and represent it in terms of the
classical functions of the operator and the density matrix,
\beq 
< \hat A > = \int dx \int d \s d \ta d \s' d \ta' \tilde A ( \s,\ta ) 
\tilde \rh ( \s',\ta' ) < x | \e ^{i \s  \hp + i \ta  \hx } 
 \e^{i \s' \hp + i  \ta' \hx }| x > .
\eeq 
Using $ [ \hp , \hx ] = -i $, $ \hx | x > = x | x > $, $\exp ( i r \hp ) 
| x > = | x - r > $, and the Campbell-Baker-Hausdorff formula,
$$ 
\e ^ {( \hat A +\hat B )} = \e^{\hat A} \e^{\hat B}  \e^{ - \frac{1}{2} 
[ \hat A , \hat B ] }, \qquad
if \qquad [ \hat A , [ \hat A,\hat B ]] = [\hat B ,  [ \hat A,\hat B ]] =0 
$$ 
one finds 
\beq 
\int dx \int d \s d \ta d \s' d \ta' \tilde\rh ( \s ,\ta ) 
\tilde A ( \s',\ta' ) 
 < x + \s +\s'  | x > \e^{ \frac{i }{2} ( \s \ta -  \s' \ta')
+ i ( \ta + \ta' ) x } ,
\eeq
and finally on account of the orthogonality of position eigenfunctions, 
\beq 
< \hat A > = 2 \pi \int d \s d \ta \tilde \rh (\s, \ta) \tilde A ( -\s , -\ta ) 
.\eeq
That reads, in terms of the classical counterparts of $\tilde A$ and the 
density matrix,
\beq 
< \hat A > = \int dp \int dx \rh ( p,x ) A ( p,x )  .
\eeq 
We conclude:  A quantum-mechanical expectation value can be written in a 
form analogous to classical mechanics where the classical counterparts of 
the operator and the density matrix are given by the Weyl correspondence 
principle. The classical density operator $\rh ( p,x )$ is commonly 
called Wigner function.  

It should be pointed out that the correspondence rule implies 
a symmetric operator ordering prescription. Thus, 
information about ordering of operators is lost 
by the transformation. However, this is an intrinsic 
property of any correspondence rule since the classical counterparts 
always commute. In other words, generally, there exist distinct 
operators which correspond to the same classical counterpart. 
        
\subsection{Wigner functions}

Given a density matrix, one may ask how to calculate the classical 
counterpart. For that purpose,  we express the Wigner function in terms 
of the matrix element, 
\beq 
< z | \hat \rh | y > = \int d\s d \ta \tilde \rh ( \s,\ta ) < z | \e^{i \s \hp
+ i \ta \hx } | y > ,
\eeq
which evaluates, using exactly the same operator relations as above, to
\beq 
< z | \hat \rh | y > = \int d\ta \tilde \rh ( y - z , \ta ) \e^{\frac{i}{2} \ta
( y + z )}. 
\eeq 
If we let $z = x - \frac{1}{2} \s $,   $y = x + \frac{1}{2} \s $, 
and perform a Fourier transformation, then
\beq
\int d\s \e^{i \s p} < x - \frac{1}{2} \s | \hat \rh |  x + \frac{1}{2} \s >
=   \int d\s d \ta \e^{ i \s p + i \ta x}   \tilde \rh (\s,\ta ) =
\rh ( p , x ) ,
\eeq
which is commonly used as definition of Wigner functions \cite{wigner}. 
The offset $\pm \frac{1}{2} \s$ in the definition accounts for the 
off-diagonal matrix elements. They are interference terms which 
-- in contrast to the diagonal ones -- 
cannot be given a probabilistic interpretation,
but nevertheless do contribute to the classical quantities.

\section{Representation in terms of creation and annihilation operators}

So far, we worked in position (momentum) representation of quantum mechanics.
In order to find a link to field theory, it turns out to be useful to go over 
to a Fock-space representation. To further develop our theory, we will need 
some definitions and relations. 
 
\subsection{Definitions and Relations}

We introduce $a,\, a^\dagger $ related to $\hat p , \, \hat x$
by
$a = \sqrt{\omega/2 } \hat x + i \hat p / \sqrt{2 \omega }$  
from which we get  
$\exp ({i \s \hat p + i \ta \hat x } )= \exp ( {\a a^\dagger - \a^* a })$
with $\a = ( i \ta - \omega \s ) / \sqrt{2 \omega }  $.
The displacement operator
\beq   \hD(\a ) = \e^{ \a \ad - \as a } = \e^{-\frac{1}{2} |\a |^2 } 
\e^{\a \ad } \e^{-\as a } = \e^{\frac{1}{2} |\a |^2 }
\e^{-\as a } \e^{\a \ad }
\eeq
has the following properties \cit{weylco}.
\begin{itemize}
\item
Unitarity:
$$ \hD^\dagger ( \a ) \hD(\a ) = \1 ,\quad \hD^\dagger (\a ) = \hD(-\a ) .$$
\item
Shift-operator:
$$ \hD^{-1}(\a )  a \hD(\a ) = a + \a , \quad   \hD^{-1}(\a )\ad  \hD(\a ) 
= \ad + \as .$$
\item The states 
$$ |\a > = \hD(\a ) | 0 > = \e^{- \frac{1}{2} |\a |^2 } 
\sum_n \frac{a^n}{\sqrt{n! }} | n > $$
are over-complete.
\item Product:
$$ < \beta | \a >=\e^{-\frac{1}{2} ( |\a |^2 + | \beta |^2 )+\a \beta^*}. $$
\item Identity: 
$$\frac{1}{\pi } \int d^2\a |\a > < \a | = \1 .$$
\item Trace:
$$ tr \hat A = \frac{1}{\pi } \int d^2 \a < \a | \hat A | \a > .$$
\item Multiplication property:
$$ tr ( \hD( \a ) \hD ( \a' ) ) = \pi \delta^2 ( \a + \a' ) .$$ 
\item Operator Fourier-decomposition:
$$ \hat A = \int d^2 \a \tilde A ( \a ) \e^ {\a \ad - \as a }  
= \int d^2 \a \tilde A ( \a) \hD (\a ).$$
\end{itemize}
On account of this and the pre-last relation, an observable amounts to
the convolution of the Fourier-components of density matrix and operator,
\beq
tr  ( \hat \rh \hat A ) = \int d^2 \a \int d^2 \a'  \tilde A(\a ) 
\tilde \rh ( \a' ) tr( \hD( \a ) \hD ( \a' ) ) = 
\int d^2 \a   \tilde A(\a ) \tilde \rh ( -\a ) .
\eeq 
The Fourier components are calculated using
\beq  \tilde A ( \a ) = tr \left( \hat A D( -\a  )  \right).
\eeq  

\subsection{Time evolution}

Time evolution is given by the von~Neumann equation,
$i \dot {\hat \rh} = [ H , \hat \rh ] $,
i.e. for $\tilde \rh ( \a ) = tr \left( \hat \rh D(-\a ) \right) $, 
\beq 
\dot { \tilde \rh }(\a )   = i\, tr \left( \hat \rh \, [ H,D (-\a ) ] 
\right) . \lab{weight}
\eeq   

\subsubsection{Example: Free motion}
 
One finds 
$$ [D(-\a ) , H ] = \om ( \a \ad + \as a + \a \as ) D ( \a ) .$$
Weightened values of creation and annihilation operators in (\ref{weight}) can be written as partial derivatives,
\beq 
tr \left( \hat \rh a  D (-\a ) \right) = \frac{ \9 \tilde \rh (\a ) }{\9 \as }
 - \frac{1}{2} \a \tilde \rh (\a ) , \qquad 
tr \left( \hat \rh \ad  D (-\a ) \right) = - \frac{ \9 \tilde \rh (\a )}{\9 \a}
+ \frac{1}{2} \as \tilde \rh (\a )  \eeq 
and the equation of motion turns into a partial differential equation
\beq 
i \dot {\tilde\rh} = \om \left( \as \frac{ \9 \tilde\rh (\a ) }{\9 \as }
- \a  \frac{ \9 \tilde\rh (\a ) }{\9 \a } \right) .
\eeq
Its solution is 
$\rh ( \a \e^{-i \om t } , \as \e^{i \om t } ) $
where we imposed the initial condition 
$\left. \tilde \rh \right|_{t=0} = \tilde\rh (\a )$.

\section{Field theory}
In order to generalize the setting to field theory, 
a Lorentz invariant formulation is called for. In that context we encounter 
a fundamental problem. A Lorentz invariant definition of an inner product
is based on a mass-shell condition. A priori, the latter -- if it exists at 
all -- is not known in an interacting theory. However, from a perturbative 
point of view, it suffices to 
know the inner invariant product of the non-interacting theory. 
This product does not pre-assume any deeper dynamic property of the 
full theory, but merely is a necessary starting point of a well defined 
perturbative expansion, which simply absorbs free motion into the 
interaction picture representation.  

\subsection{Lorentz invariant product in the free case}

Particles of mass $m$ define two hyperboloids $k_0=\pm \om_k$ with
energy $\om_k = \sqrt{\bk^2 + m^2 }$. 
Real on-shell functions on space-time can be Fourier decomposed as
\beq
T ( t,\bx )  = \frac{1}{(2 \pi )^{3/2} } \int ( d\mu^+ ( k) 
+  d\mu^- (k) ) e^{-i kx } \ta ( k ) 
\eeq
where  $d\mu^\pm ( k ) = d^4 k\de ( k^2 - m^2 ) \theta ( \pm k_0 ) $
are the measure components invariant under orthochronical homogeneous 
Lorentz transformations  $\in \mathcal  L^{\uparrow}_+ $.
An invariant inner product for real fields is given by 
\beq
(T, \Sigma ) = \int ( d\mu^+ (k ) +   d\mu^- (k ) )
 \ta^* (k) \s (k )  
= \int d\mu^+ ( k )  ( \ta^* (k) \s (k )
+ \ta (k ) \s^* (k ) ) 
\eeq
which reads in position space 
\beq (T, \Sigma ) = i \int d^3x \left( T^* (t,\bx ) \9_t  \Sigma (t,\bx ) - 
(\9_t T^* (t,\bx ) ) \Sigma ( t,\bx ) \right) .
\eeq
The direction $\9_t$ characterizes the space-like hyper-surface on 
which the quantisation is performed. The inner product is constant along 
that direction. 
 
The field operator in interaction picture representation  
\beq \Phi_I(t,x) =\frac{1}{(2 \pi)^{3/2} } \int d\mu^+(k) 
( b (\bk) e^{- i k x } + 
b^\dag ( \bk ) e^{i k x } )
\eeq
decomposed into creation and annihilation operators with commutator
\beq 
[ b( \bk ),  b^\dag ( \bk ) ] = 2 \om_k \de^3 ( \bk - \bk' ) 
\eeq
evolves with the free Hamiltonian
\beq 
\dd H_0 \dd = \int d\mu^+ (k)  \om_k b^\dag (\bk ) b ( \bk ) . 
\eeq
The rescaled definition  of creation and annihilation operators 
$a(\bk ) = \sqrt{2 \om_k } b ( \bk ) $ used in the previous section 
in the zero-dimensional case, and commonly advocated to in textbooks,
was introduced and discussed by Newton and Wigner \cit{NW}. One may regard
the corresponding wave function as a Lorentz-frame dependent 
probability amplitude since their scalar product in position space 
takes the same form as in the Schr\"odinger theory.

The exponent of the shift operator evaluates to
\beq 
i ( \Sigma , \Pi_I ) + i (T , \Phi_I ) =
\int d\mu^+ (k) ( \be ( \bk ) b^\dag ( \bk ) - \be^* ( \bk ) b( \bk ) ) ,
\qquad  \be( \bk ) = i \ta (\bk ) - \om_k  \s ( \bk ) ,
\eeq
where we took into account that $\Pi_I = \dot \Phi_I$.
For two reasons, the definition of $\be $ is chosen to 
differ by a factor $ \sqrt{2 \om_k } $ from 
the analog relation for $\a$. 
The exponent formally has a Lorentz invariant form and the 
properties and relations derived in the previous section can be 
carried over to field theory straightforwardly. 
In particular, the coherent states chosen as basis are created by acting with 
the shift operator functional in interaction picture representation
\beq  
D_I[\be ] = \exp \left\{ {\int d\mu^+ (k) ( \be ( \bk ) b^\dag ( \bk ) -
\be^* ( \bk ) b( \bk ) ) }  \right\}
\eeq 
on the vacuum. Note that only the introduction of a Lorentz
invariant product renders the exponent time independent, which will
turn out crucial in what follows.

\subsection{Full dynamics}

It is useful to split off free (unperturbed) motion 
by going to the interaction picture representation of $\hat \rho $
which obeys the evolution equation 
\beq 
i \dot {\hat \rho_I} (t )  = [ W_I (t) , \hat \rho_I ( t ) ].
\eeq  
We introduce a Wigner functional decomposition
\beq 
\hat \rho_I (t) = \int   {\mathcal D}^2 \be \rho_I [\be ] (t)D_I [\be ]
\eeq
and its inverse, the Wigner transform of the interacting 
density matrix,
\beq 
\rho_I [ \be  ] (t) = \tr \left( \hat \rh_I (t)  D_I [-\be ] \right) ,
\eeq
which is subject to the equation of motion
\beq 
\dot \rho_I [\be ] (t) = 
i \tr \left( \hat \rho_I (t) \left[ W_I (t) ,D_I[ -\be ] \right] \right).
\lab{eom}
\eeq
The evaluation of time dependent weightened operator expectation values,
\beq 
\tr \left( \hat \rho (t) \hat A \right)= 
\tr \left( \hat \rho_I (t) \hat A_I (t) \right) =
\int {\mathcal D}^2 \be \rho_I [-\be ](t) A_I [ \be ] 
\eeq
also calls for calculating the interacting operator Fourier components  
$ A_I [ \be ] $,  
\beq 
A_I [ \be  ] (t) = \tr \left( \hat A_I (t)  D_I [-\be ] \right) .
\eeq

\subsection{Equation of motion}

Let us examine the commutator  $\left[ W_I (t) ,D[ -\be ] \right] $ 
at the r.h.s.\ of the equation of motion more closely. 
We suppose that the interaction is 
a polynomial in the fields and momenta, to wit, also a polynomial 
in creation and annihilation parts.  
Since $D $ acts as a shift operator, (We drop the index I in what follows.)
\beq 
D[ -\be ] b( \bk ) = ( b (\bk ) + \be (\bk ) ) D[ -\be ],\quad 
D[ -\be ] b^\dag ( \bk ) = ( b^\dag (\bk ) + \be^* (\bk ) ) D[ -\be ]
\eeq
we get for the commutator 
\beq \left[ W[\Phi ] ,D[ -\be ] \right] = 
( W[\Phi ] -  W[ \Phi + \Psi ] ) D[ -\be ]
\eeq 
with 
\beq 
\Psi (t,\bx ) =  \frac{1}{(2 \pi)^{3/2} } \int d\mu^+ (k) 
( \be (\bk ) e^{ - i k x } + 
\be^* ( \bk ) e^{i k x} ) .
\eeq
In order to draw the operator-valued commutator 
out of the trace in Eq.\ (\ref{eom}), it is useful to represent 
$b$ and $b^\dag$ by suitably chosen functional derivatives acting on $D$,
\bea
b( \bk )  D[ -\be ] = \left( 2 \om_k  \frac{\de}{\de \be^* ( \bk ) } - 
\frac{1}{2} \be (\bk ) \right)  D[-\be ]  ,\nn \\  
b^\dag ( \bk )  D[ -\be ] = \left( - 2 \om_k  \frac{\de}{\de \be ( \bk ) } - 
\frac{1}{2} \be^* (\bk ) \right)   D[-\be ] \lab{dif}.
\eea
Products of operators $b$ and $b^\dag$ can be represented by replacing them
with their partial derivative counterparts in inverse order.
The terms with the factor $\frac{1}{2}$ in front originate from the 
fundamental commutator and are of relative order $\hbar$. The 'classical' 
field operator 
\beq 
\bP (t,\bx ) = \frac{1}{(2 \pi)^{3/2} } \int d\mu^+ (k)  2 \om_k
\left(  e^{ - i k x }  \frac{\de}{\de \be^* ( \bk ) }
- e^{i k x} \frac{\de}{\de \be ( \bk ) } \right) 
\eeq 
commutes with the displacement $\Psi (t,\bx ) $ which permits us to 
write the commutator as 
\beq
 ( W[\bP - \frac{1}{2} \Psi  ] -  W[ \bP+ \frac{1}{2} \Psi ] ) D[ -\be ].
\eeq 
The equation of motion (\ref{eom}) for the interacting part of the density
matrix thus assumes the form of a Liouville
equation
\beq 
\dot \rho [\be ] (t) = i L[\be ](t) \rho [\be ] (t) ,\quad  
L[\be ](t) =   ( W[\bP - \frac{1}{2} \Psi  ] -  W[ \bP+ \frac{1}{2} \Psi ] ).
\lab{eom3}
\eeq
A formal solution can be constructed by standard methods, 
\beq
\rho [\be ] (t)=\ct \exp \left\{ i \int_{t_0}^{t} dt' L[\be ](t') \right\} 
\rho [\be ] (t_0 ) \lab{eom2} 
\eeq 
using the time ordering operator $\ct$. This equation contains the full
dynamical information on the Wigner transform of the density operator. 
The equations of motion of the c-number counterpart of $\hat \rho $
can be obtained by a functional integration with the classical field and 
momentum in the displacement operator similar to Eq.\ (\ref{toclass}).

\subsection{Equilibrium distribution}

In order to illustrate the formalism derived, we consider a 
non-interacting system at thermal equilibrium  
in which interaction is switched on at some initial time.   
In general, canonical ensembles in thermal equilibrium with temperature 
$ T $ are characterized by the density matrix
$\hat \rho_{eq} = \e^{ - H/T }.$
Due to standard arguments in perturbation theory, this density matrix
can be rewritten as 
\beq
\hat \rho_{eq} = \e^{-  H_0 /T } U_I ( t_0 - i /T , t_0 ) ,
\qquad U_I ( t_0 - i/T , t_0 )  = 
\ct \exp \left\{ -i \int_{t_0}^{ t_0 - i/T } dt' W_I (t' )\right\}.
\eeq
The corresponding Wigner function in interaction picture
is given by 
\beq 
\rho_{eq} [ \be ](t) = \tr \left(\e^{- H_0 /T } 
U_I ( t_0 - i/T , t_0 ) D[ -\be ] \right) .
\eeq
Let us calculate the non-interacting equilibrium Wigner function defined by 
\beq
 \rh^0_{eq}  [\be ] = \tr \left( \e^{-H_0 /T } D[ -\be ] \right)
\eeq
which we will assume to be the initial preparation of the system.
For that purpose, we consider the non-interacting 
part with one creation operator
inserted between the statistical factor and the shift operator. 
Commutating $b^\dag$ to the right, using the cyclicity of the trace and the 
commutator with $H_0$ we find 
\beq
\tr \left( \e^{-H_0 /T} b^\dag (\bk ) D[ -\be ] \right)= 
-\be^*(\bk )\rh^0_{eq} [\be ] + \e^{ \om(\bk)/T } \tr \left(\e^{- H_0 /T} 
b^\dag (\bk ) D[ -\be ] \right) .
\eeq
On the other hand one gets  $b^\dag$ in the trace by acting with the 
differential operator of Eq. (\ref{dif}) on $D$,
\beq
\left( - 2 \om_k  \frac{\de}{\de \be ( \bk ) } - 
\frac{1}{2} \be^* (\bk ) \right) \rh^0_{eq}  [\be ] = 
\tr \left( \e^{-H_0 /T} b^\dag (\bk ) D[ -\be ] \right).
\eeq 
Combining this property with the previous equation and functionally 
integrating with respect to $\be ,$ one finds,
up to a normalisation constant,
\beq
\rh^0_{eq} [\be ] = \exp \left\{ \int d\mu^+ (k )  
\left( \frac{1}{2} - n^- (\bk ) \right) 
\be ( \bk )  \be^* ( \bk ) \right\} \lab{z0}
\eeq
with $n^\pm (\bk )  = (1 - \exp ( \pm \om_k /T  ))^{-1} .$ 

\subsection{Non-interacting partition function as initial condition} 

We consider the non-interacting partition function (\ref{z0}) as initial
condition in the time evolution (\ref{eom2}) and expand the exponent in 
powers of the coupling constant.
To be specific, we consider a quartic interaction term in the Hamiltonian, 
\beq
W[\Phi] = \frac{\la}{4 ! } \int d^3 x \Phi^4 (x)  .
\eeq
We act with the differences of Eq.\ (\ref{eom3}) on $\rh^0_{eq} [\be ] $ 
and obtain
\beq 
(\bP (x) + \frac{1}{2} \Psi (x) )  \rh^0_{eq} [\be ]  = 
\G ( x) \rh^0_{eq} [\be ] ,\qquad 
(\bP (x) - \frac{1}{2} \Psi (x) )  \rh^0_{eq} [\be ] = 
- \G^* ( x) \rh^0_{eq}  [\be ]
\eeq
with 
\beq 
\G (x) = \frac{1}{(2\pi )^{3/2}} \int d\mu^+ (k)
\left( \be(\bk) n^+ (\bk ) \e^{-i k x }+ 
\be^*(\bk) n^- (\bk ) \e^{i k x } \right).
\eeq	
Green functions arise in the commutator of the differences with $\G$,
\beq 
\left[ \bP (x) \pm \frac{1}{2} \Psi (x) ,\G (y) \right] = G (x-y ) , \qquad 
\left[  \bP (x) \pm \frac{1}{2} \Psi (x) ,\G^* (y) \right]  =  -G^* (x-y ), 
\eeq
with 
\beq
G(x-y) = \frac{1}{(2 \pi )^3 } \int d\mu^+(k) ( e^{-i k (x-y ) } n^- (\bk)-
e^{i k (x-y ) }  n^+ (\bk)).
\eeq  
The first order term in $\la$ is found to read 
\beq 
L[\be](x_0)\rh^0_{eq}[\be] = \frac{\la}{4!}\int d^3 
x\cW[\be](x)\rh^0_{eq} [\be ] \eeq
with 
\beq
\cW ( x ) =  6 G(0) ( \G^* (x)^2 -  \G (x)^2 )
+ ( \G^*(x) ^4 - \G(x) ^4 ) .
\eeq 
For higher orders, it turns out useful to introduce a graphical 
representation. 
We denote a factor $\G$, $\G^* $ by an external leg. Green functions 
and their complex conjugate correspond to internal lines with an without star. 
Each power of the interaction term gives rise to a vertex.

\begin{figure}
\begin{center} 
\epsfbox{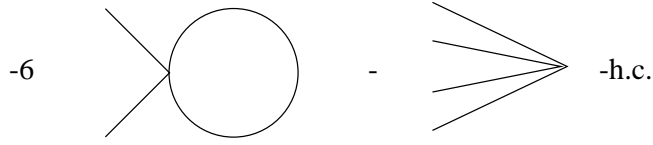}  
\caption{First order contribution.}
\lab{figfig}
\end{center}
\end{figure}

In that way, the second order contribution 
\beq
L[ \be](x_0)L[\be ](y_0) \rh^0_{eq}[\be ] =\frac{\la^2}{(4!)^2} \int d^3 x 
\int d^3 y \cW [\be ](x,y) \rh^0_{eq}  [\be ]     
\eeq 
can be represented by the graphs 
(\ref{fig2}) trough (\ref{fig8}),
where legs on the l.h.s.\ and r.h.s.\ 
correspond to the arguments $x$ and $y$ of $\G$. $\cW [\be ](x,y)$ involves 
contributions with two, four, six and eight external legs and is real.

\begin{figure}
\begin{center} 
\epsfbox{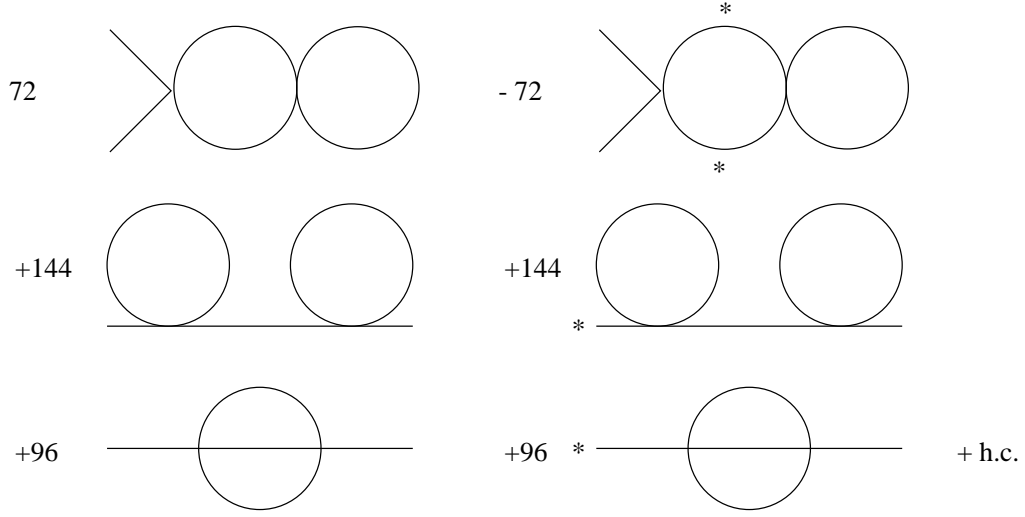}  
\caption{Second order with two external legs.}
\lab{fig2}
\end{center}
\end{figure} 

\begin{figure}
\begin{center} 
\epsfbox{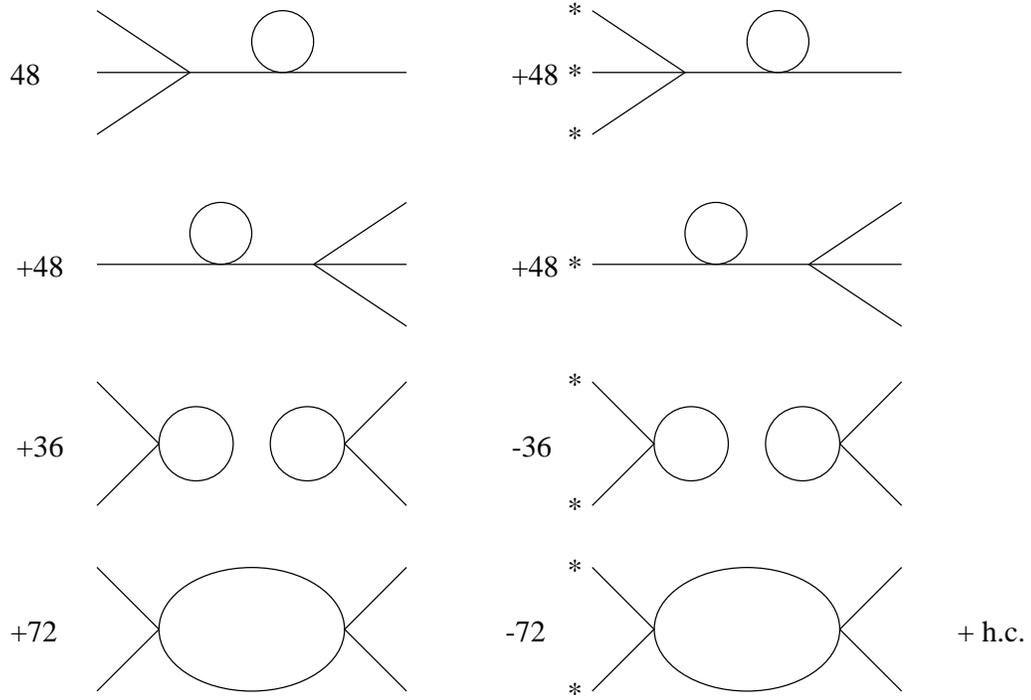}  
\caption{Second order with four external legs.}
\end{center} 
\end{figure} 

\begin{figure}
\begin{center}
\epsfbox{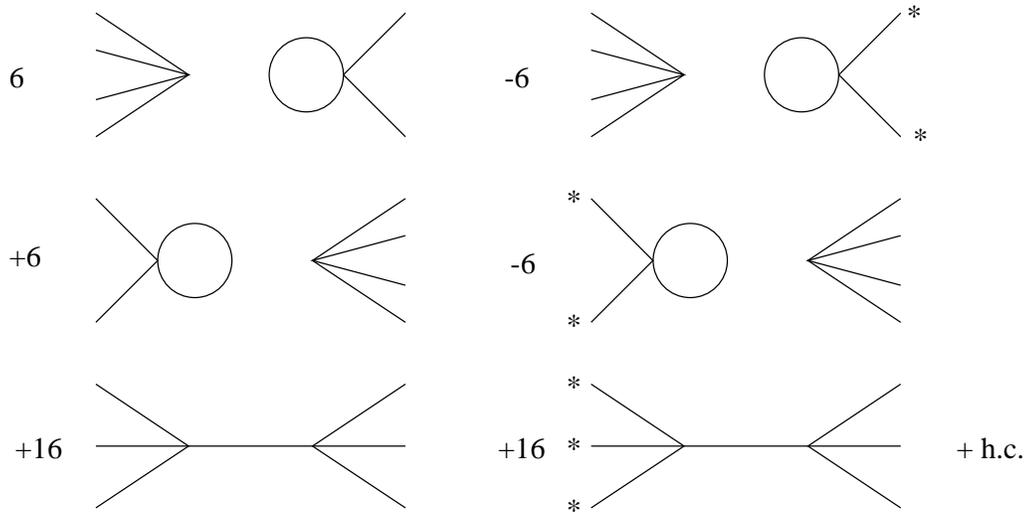}  
\caption{Second order with six external legs.}
\end{center}
\end{figure} 

\begin{figure}
\begin{center}
\epsfbox{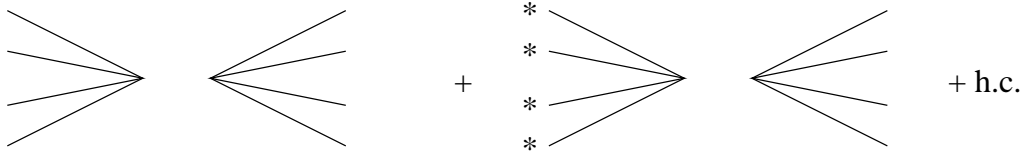}  
\caption{Second order with eight external legs.}
\lab{fig8}
\end{center}
\end{figure} 

The time dependent density matrix in $\be$-presentation with the given 
initial condition is thus given by (We take $t_0=0$.)
\bea
\rho [\be ] (t )  &= &  \left( 1 + \frac{i\la}{4!} \int_{0}^t dx_0 
\int d^3x \cW (x)[\be ]  -  \right. \nn \\ 
&& \left. -  \frac{\la^2}{ (4!)^2}
\int_{0}^t dx_0 \int_{0}^{x_0} dy_0 \int d^3x \int d^3y \cW (x,y) [\be]
\right) \rh^0_{eq}  [\be]       . \lab{dens}
\eea
These expressions contain all information about any operator 
expectation value to order $\la^2$. 

\subsubsection{Particle density}

The extraction of the particle density is achieved as follows.
We act with the bilocal particle number operator $n(\bk_1,\bk_2) = 
b^\dag (\bk_1 ) b(\bk_2) $ in 
$\be$-representation on the time dependent density matrix (\ref{dens}) 
and set $\be=0$ afterwards. In particular,  commuting 
\beq 
n[\be ] (\bk_1,\bk_2) =  
\left( 2 \om_{k_2}   \frac{\de}{\de \be^* ( \bk_2 ) } - 
\frac{1}{2} \be (\bk_2 ) \right)    
\left( - 2 \om_{k_1}  \frac{\de}{\de \be ( \bk_1 ) } - 
\frac{1}{2} \be^* (\bk_1 ) \right)   
\eeq 
to the r.h.s. using 
\bea
\left[ b^\dag[\be](\bk) ,\G (x) \right] =
- \frac{ n^+(\bk) }{(2 \pi )^{3/2} } e^{-i k x }, \quad 
\left[ b [\be](\bk) ,\G (x) \right] =  
\frac{ n^-(\bk) }{(2 \pi )^{3/2} } e^{i k x } ,
\nn \\
\left[ b^\dag[\be](\bk) ,\G^* (x) \right] = 
- \frac{ n^-(\bk) }{(2 \pi )^{3/2} } e^{-i k x }, \quad
\left[ b[\be](\bk) ,\G^* (x) \right] =  
\frac{ n^+(\bk) }{(2 \pi )^{3/2} } e^{i k x }  
\eea
essentially replaces two external legs by plane waves weightened with 
distribution functions $n^\pm (\bk ) $. 
Thus, after having set  $\be =0$, the set of graphs that can 
contribute are those with two external legs only.
Moreover, explicit calculation shows that the diagrams 
in Fig.\ (1) vanish and the first two graphs of Fig.\ (\ref{fig2})
cancel after integration over $x$. 
The zeroth order term gives ($\om_1 \equiv \om_{\bk_1} $)
\beq 
- 2 \de^3 ( \bk_1 - \bk_2 )  \om_1 n^+ ( \bk_1 ) 
\eeq 
and the third and fourth diagram in Fig.\ (\ref{fig2}) contribute, after 
integration over $\bx,\by$ and $x_0,y_0$ 
\beq
\la^2 \de^3 (\bk_1 - \bk_2 ) \frac{G^2(0) }{(2 \om_1)^3  } 
( n^+ (\bk_1 ) - n^- ( \bk_1 ) ) \sin^2 (\om_1 t ).
\lab{harm} \eeq
The sunset-diagrams in  Fig.\ (\ref{fig2}) are the ones far most complicated. 
They contribute $( \ta = x_0 - y_0 )$
$$
\la^2  \frac{\de^3(\bk_1- \bk_2 ) }{6(2 \pi)^9} 
\int_0^t dx_0 \int_0^{x_0} d y_0  ( \e^{-i \om_1  \ta  } 
n^-(\bk_1) + \e^{i \om_1 \ta  } n^+(\bk_1) ) 
\int d^3 x \e^{ i \bk_1 \bx } \ccl^3 ( x, \ta ) 
+ \mbox{h.c.} ,
$$
\beq 
\ccl (x,\ta ) = \int \frac{d^3 p}{2 \om_p } ( \e^{-i \om_p \ta + i\bp \bx}
n^- ( \bp ) -  \e^{i \om_p \ta - i\bp \bx} n^+ ( \bp ) ) . 
\lab{57} 
\eeq
Let us discuss  the contribution (\ref{harm}) in more detail. 
It exhibits a harmonic
time dependence and does not decrease with time as one may expect for
the particle number density which presumably  approaches an equilibrium
value. The situation is even worse for off-diagonal 
operator expectation-values. One finds for the contribution from the third 
and fourth diagram in  Fig.\ (\ref{fig2}) to the operator $b(\bk_1 ) b(\bk_2) $
\beq 
\la^2 \de^3(\bk_1 + \bk_2 ) \frac{ G^2(0) }{(2 \om_1)^3 }
\left(  \sin^2 \om_1 t + i ( t \om_1 - \sin \om_1 t \cos \om_1 t ) 
\right) ( n^+ ( \bk_1 ) - n^- (\bk_1 ) ) ,
\eeq 
which even contains terms linear in time. This 
indicates a breakdown of the expansion of the time ordered exponent in the 
evolution (\ref{eom2}). A crude estimate of the domain of validity of 
the expansion
performed can be given by the requirement that the exponent 
remains smaller than unity, i.e.\ for times $\la t \ll 1$.

\subsection{Resummation} 
As well known in equilibrium finite
temperature field theory, consistency in the infrared regime 
requires a resummation to be carried out to reorganize the 
perturbative series \cit{equil}. As anticipated in the example
given, for non-equilibrium density distributions 
the necessity of resummation also arises for dynamical reasons. 
Here, resummation boils down to add a mass square $M^2$ to the 
original Hamiltonian and compensate it by a two-vertex
\beq  
- \frac{1}{2} \int d^3 x M^2 \Phi^2 (x)
\eeq 
in the interaction part. This term changes the Liouville-operator 
in the equations of motion, which gives rise to a number of additional 
diagrams. Those are such that each graph with closed loops $\propto G(0) $
gets a mass counterpart in a way that they compensate for the choice 
\beq 
M^2 = \frac{\la}{2}  G(0), \lab{rescond}
\eeq 
as illustrated in Fig.\ (\ref{fig6}).

\begin{figure}
\begin{center}
\epsfbox{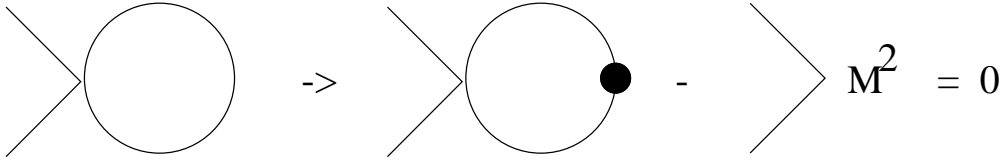}  
\caption{Resummation of mass tadpoles}
\lab{fig6}
\end{center}
\end{figure} 

The resummed mass $m^2 + M^2 $ appears now hidden in the Lorentz-covariant
volume element $d\mu^+ $ and the corresponding energy eigenvalues
$\sqrt{ m^2 + M^2 + k^2 } $ both appearing in $ G(0)$. In that sense, 
Eq.\ (\ref{rescond}) is a consistency condition for the mass counter term.  
It is interesting to observe that not the first order dynamical contributions 
to the particle number, which vanish, enforce the resummation, but the time 
evolution of the second order terms (\ref{harm}) necessitate the 
reorganisation of the perturbative series. The redefined theory gets 
its second order contributions (\ref{57}) solely from the non-local 
sunset-graphs.

\section{General initial conditions} 

We generalize the particular density matrix 
considered in the previous section by a Wigner density functional 
of the connected graphs expressed in a functional 
power series in $\be$ (See Appendix.), i.e.\
\beq 
\log \rho [\be ](t) = \sum_{n=1}^{\infty } \sum_{\lb \s \rb}
\frac{1}{n!}
\left( \prod_{i=1}^{n} \int d\mu^+ (k_i) \right)
f_n^{\lb \s \rb } ( t;\bk_1 , \ldots ,\bk_n )
\be^{\s_1} (\bk_1  ) \ldots \be^{\s_n} ( \bk_n ) , \lab{logdef}
\eeq 
where elements of the set 
$\lb \s \rb= \lb \s_1,\ldots,\s_n\rb $ 
can assume values $+$ and $-$ which refer to starred an unstarred $\be$'s
respectively. The correlators $f_n$ may also explicitly depend on time.
For a given order in $\be$, not all functions $f_n$
are independent. They have to 
be symmetric with respect to pairwise exchange of 
two momenta, both corresponding to starred ($\s=+$) or   
unstarred ($\s=-$) pairs of $\be$'s. Furthermore,
since physical density matrices have to be hermitian,
the relation $D[\be] = D[-\be]^\dag $ implies invariance 
of $\rho[\be] $ with respect to simultaneous hermitian
conjugation and sign flip of $\be$. This implies for 
the correlators $f$
\beq  { f_n^{\lb \s \rb }}^\dag (t; \bk_1 , \ldots ,\bk_n ) =
(-)^n f_n^{\lb \s^* \rb } ( t;\bk_1 , \ldots ,\bk_n )
\eeq 
The number of independent correlators is $(n+2)/2$ for $n$ even
and $(n+1)/2 $ for $n$ odd.

Let us study the r.h.s.\ of the equation of motion (\ref{eom3}).
Acting with the Liouville operator on $\rho[\be] = \exp (\log\rho[\be])$ 
gives 
\bea
L[\be] \rho[\be] = - \frac{\la}{6} \int d^3x \Psi(x)\left( 
3 [\bP(x),[\bP(x) ,\log \rho ]] [\bP(x) ,\log \rho ] +\hspace*{2cm}
\nn \right. \\
+ ( [\bP(x) ,\log \rho ])^3 +
 \left. [\bP (x),[\bP(x), [ \bP(x),\log \rho]]] +
\frac{1}{4} \Psi^2(x) [\bP (x),\log \rho ] \right) \rho .
\lab{lrho}   \eea
The first few terms may best be studied by introducing a graphical
notation. We represent $\log \rho$ by the graphs depicted in 
Fig.\ (\ref{lnrho}). 

\begin{figure}
\begin{center} 
\epsfbox{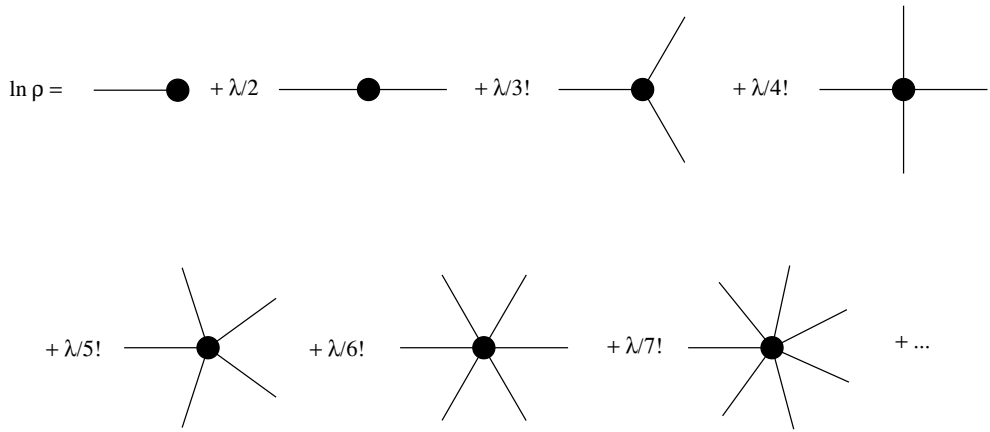}  
\caption{$n$-th order term in $\log \rho$.}
\lab{lnrho}
\end{center}
\end{figure}  

The blob in the middle denotes the correlators $f$, and each
external leg a factor containing $\be$ or $\be^*$. We do not 
explicitly split the sum over $\lb \s \rb$ into their summands 
 -- each graph is understood as the sum over all independent 
$f_n$'s for a fixed order $n$. 
Acting with $\bP(x)$ on an external leg cuts it off and replaces 
it by a plane wave. We denote cut legs by a cross.
The factor $\Psi$ is represented by a cut leg without blob
since it corresponds to a unit cut one-point correlator.
The integration over space graphically joins all crossed legs in the 
point $x$. The time derivative $i \9_t$ in the equations of motion is 
symbolized by a dot. The motion of the system to order $\be^4$ 
is depicted in Figs.\ (\ref{eom1f}-\ref{eom4f}). 

\begin{figure}
\begin{center}
\epsfbox{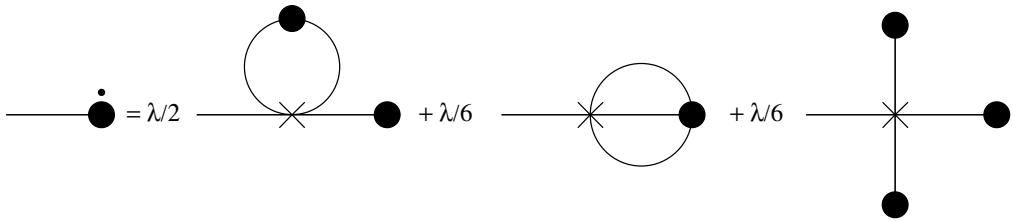}  
\caption{First stage of the hierarchy of the equations of motion.}
\lab{eom1f}
\end{center} 
\end{figure}

\begin{figure}
\begin{center}
\epsfbox{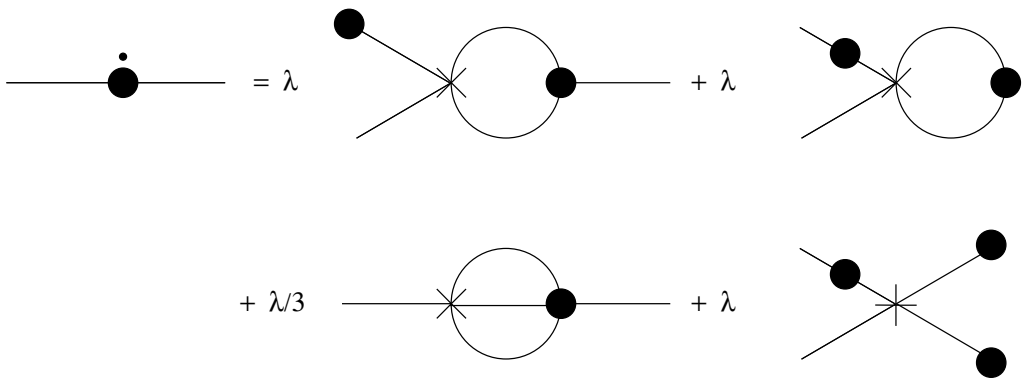}  
\caption{Second stage of the hierarchy of the equations of motion.}
\lab{eom2f}
\end{center} 
\end{figure}     

\begin{figure}
\begin{center}
\epsfbox{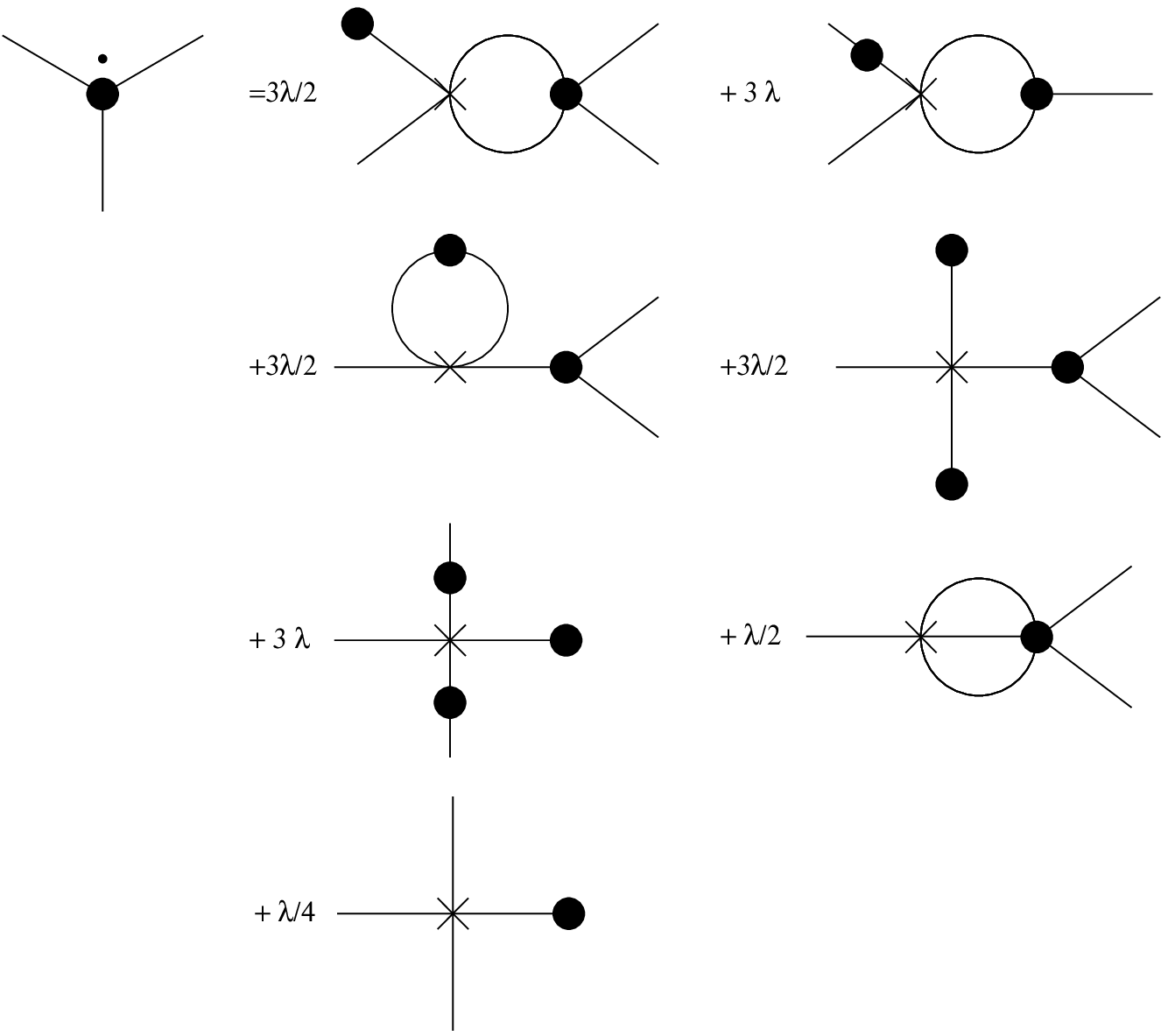}  
\caption{Third stage of the hierarchy of the equations of motion.}
\lab{eom3f} 
\end{center} 
\end{figure}

\begin{figure}
\begin{center} 
\epsfbox{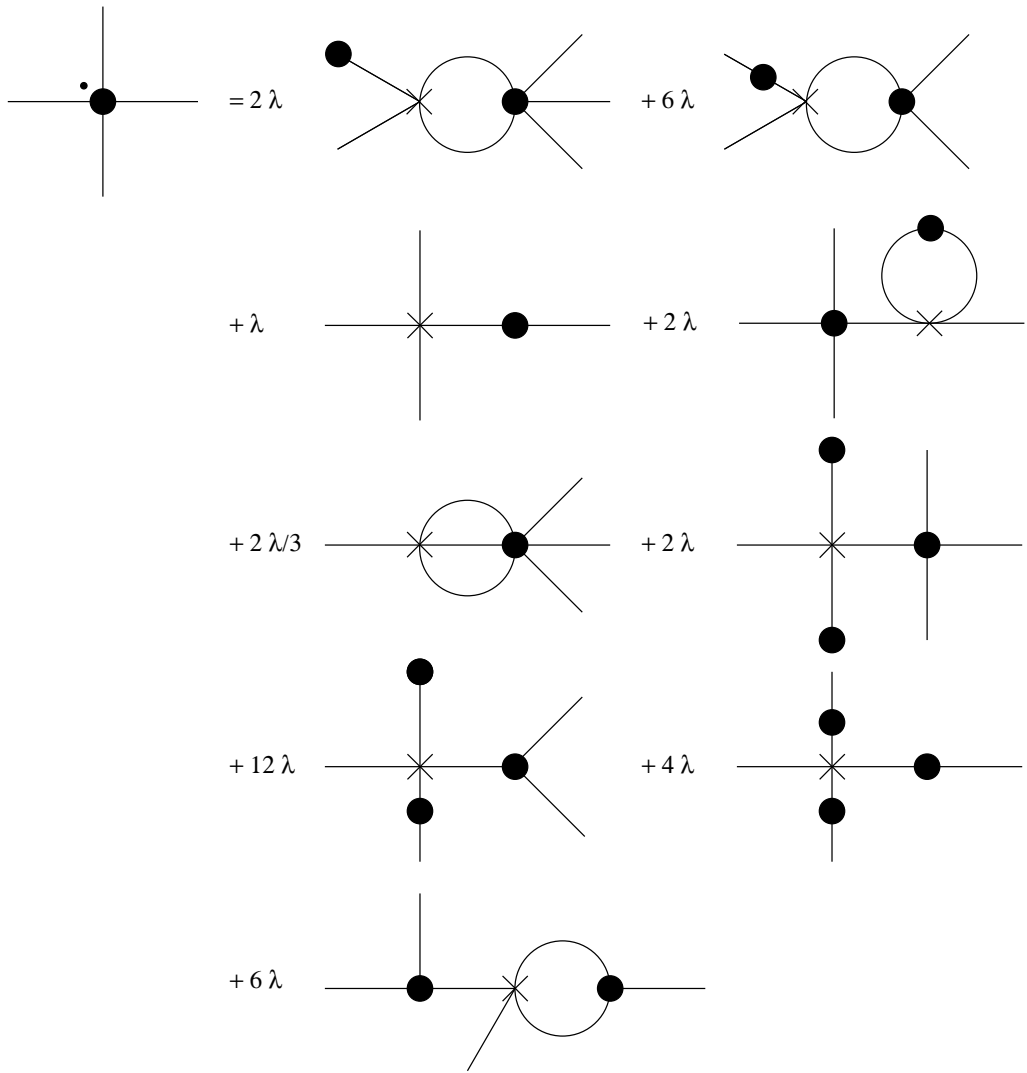}  
\caption{Fourth stage of the hierarchy of the equations of motion.}
\lab{eom4f}
\end{center}
\end{figure}

Let us clarify the magnitude  of these graphs in orders of $\hbar$.
The relevant quantities have the following dimension. 
\vspace{1em}

\begin{tabular}{|l|c|c|c|c|c|c|c|}
\hline
object & $L[\be] $ & $\bP$ & $\Psi$ & $\be$  & $\de / \de \be$ & $f_n$ \\  
\hline
order in $\hbar$ & $1$      & $1/2$ & $3/2$  & $-1/2$ & $-5/2 $ & $-3n/2$ \\ 
\hline
\end{tabular}
\vspace{1em} 

One can easily check that all graphs are of the same order in $\hbar$
except for the last in Fig.\ (\ref{eom3f}) and the third at the r.h.s.\ 
in Fig.\ (\ref{eom4f}),
which are suppressed by a factor  $\hbar^2 $.
In general, all those terms originate from the last summand 
$\propto \Psi^2$ in Eq.\ (\ref{lrho}). 
A systematic expansion thus requires to keep all but
those terms of higher order. However, the structure of the 
the whole
hierarchy is such that the time derivative of correlators of order $n$ 
involves correlators of order $n+2$ and lower on the r.h.s.\ in the equations 
of motion. The $\hbar^2$ suppressed terms are of order $n-2$ in $\be$. 
In Fig.\ (\ref{eomn}), for any $n$, all graphs are shown which involve 
higher or equal order correlators at the r.h.s. They are responsible for 
the fact that cutting the hierarchy gives a non-closed system of equations.
Even worse, the hierarchy cannot be cut at some order by the condition 
$f_n=0$ for $n$ grather than some fixed number, since that would contradict the equations for 
$\dot f_n $ containing nonvanishing contributions at the r.h.s\ 
coming from lower order correlators. 
 
\begin{figure}
\begin{center}
\epsfbox{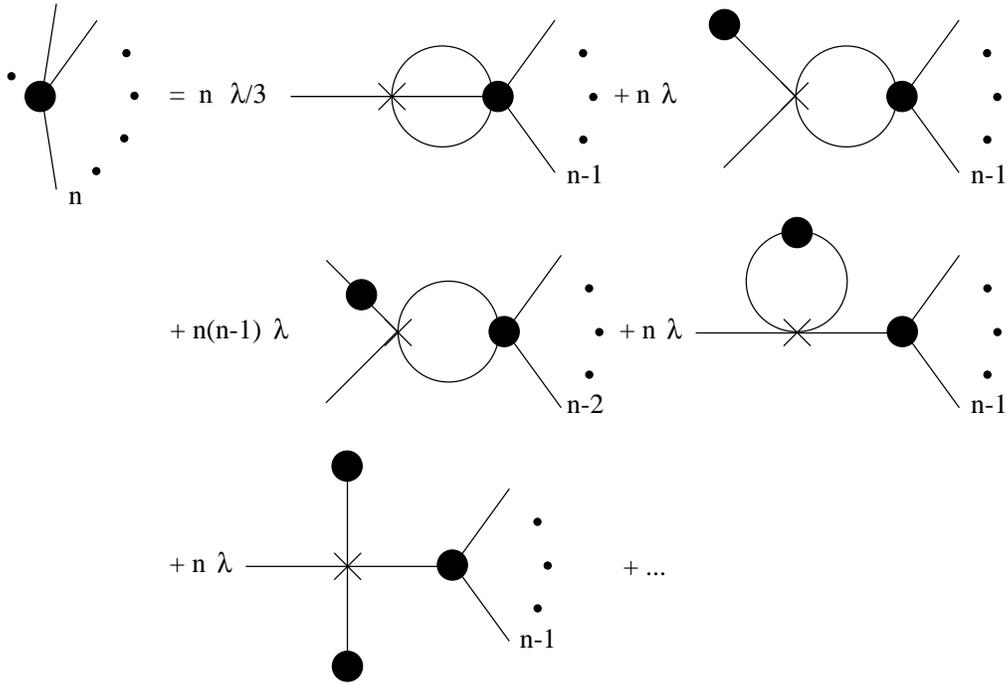}  
\caption{n-th order stage of the equations of motion.}
\lab{eomn}
\end{center}
\end{figure}      

\section{Particular configurations} 

Many systems of physical interest exhibit particular symmetries which 
imply restrictive conditions on the correlators $f_n$.   

\subsection{Spatially homogeneous systems} 

Spatially homogeneous systems are characterized by the invariance of 
operator products with respect to translation in space, to wit,
\beq 
\tr \left(\hat \Phi (t,\bx_1 ) \hat \Phi (t,\bx_2 ) 
\ldots \hat\Phi ( t,\bx_ n ) \hat\rho \right) = 
\tr \left(\hat \Phi (t,0 )\hat \Phi (t,\bx_2 - \bx_1 )  \ldots 
\hat\Phi ( \bx_ n - \bx_1 ) \hat\rho \right) . \lab{trans}
\eeq 
The corresponding expression in terms of Wigner transforms is found by 
replacing  $\Phi (t,\bx ) $ by $  \bP (t,\bx ) - 
\frac{1}{2} \Psi (t,\bx) $, the density operator by its transform  and the 
trace by a functional integral. Let us pick that term which only contains  
$ \bP (t,\bx_i )$'s, act with it on $\rho [\be ] $ and set $\be = 0 $ 
afterwards. Without loss of generality we may furthermore set $t=0$ since
time evolution commutes with spatial translations. The condition 
(\ref{trans}) implies  
\bea
\left( \prod_{i=1}^n \int d\mu^+(k_i)  \e^{ i  \bk_i \bx_i \s_i } \right)
f^{ \lb \s \rb }_n (\bk_1,\ldots , \bk_n) = \qquad\qquad\qquad\qquad\qquad
  \nn \\
\left( \prod_{i=1}^n \int d\mu^+(k_i) \right)  
\exp \left( i \sum_{i=2}^n \bk_i (\bx_i -\bx_1 ) \s_i \right) 
f^{ \lb \s \rb }_n (\bk_1,\ldots , \bk_n) .
\eea
However, this equation can only hold for correlators which are such that
the exponents at both sides are equal. That implies the particular
form
\beq 
f^{ \lb \s \rb }_n (\bk_1, \ldots ,\bk_n ) = 
f^{ \lb \s \rb ,hom}_n (\bk_1, \ldots ,\bk_n )
\del \left( \sum_{i=1}^n \s_i \bk_i \right) 
\eeq 
where the homogeneous correlator includes one redundant momentum. The 
delta-function of the $\s$-averaged momenta generalizes momentum 
conservation for spatially homogeneous systems to the correlators. 
Spatial homogeneity and isotropy is a constant of motion, to wit, 
homogeneous initial conditions evolve to homogeneous correlators.
 
\subsection{Unbroken symmetry.} 
Another physically interesting particular situation are configurations
with unbroken symmetry. They are characterized by vanishing expectation
values of the field operator, i.e.\ 
\beq 
\tr \left(\hat \Phi (t,\bx )  \hat\rho \right) = 0 .
\eeq   
Again, this condition implies  a restriction on the Wigner transforms.
The operator   $  \bP (t,\bx ) - 
\frac{1}{2} \Psi (t,\bx) $ decreases ($ \bP$) resp. increases ($\Psi$)
the number of $\be$'s in the trace by one. Comparing all terms
with equal number of $\be$'s, which have to vanish separately, 
one easily finds that all odd correlators have to vanish, 
 \beq 
f^{ \lb \s \rb ,hom}_n (\bk_1, \ldots ,\bk_n )=0, \quad n\in \N_{odd} . 
\eeq 
Closer inspection of the graphical representations given in 
Figs.\ (\ref{eom1f}-\ref{eom4f}) and their higher order counterparts 
reveals that the time derivative of odd correlators also vanishes if
they represent a system with unbroken symmetry, which is necessary 
for consistency. The symmetry 
conservation is a consequence of the invariance 
of the Hamiltonian under sign flip of the field.

\newpage

\section{Conclusion and Outlook} 
In this paper we reformulated the von~Neumann equation for the 
density matrix in terms of the classical observables of field 
correlators using the Weyl correspondence principle and the language
of Wigner functions. We found a Liouville equation for the 
Weyl-transform which expands into an infinite hierarchy of equations
of which we showed that they cannot be cut consistently at a finite 
order. Time evolution is 
known exactly, and can be -- at least perturbatively -- be integrated 
to later times. Further modelling of physical scenarios,
however, includes additional statistical assumptions entering
in realistic initial conditions on the density matrix. Based on the 
present work, investigations in that direction are currently under 
progress.

\bigskip \bigskip \bigskip
\noindent Acknowledgement: I thank M.\ LeBellac and F.\ Guerin for 
fruitful discussions and the {\em Institut Non Lin\'eaire de Nice}
for their kind hospitality. This work was supported by the  
{\em \"Osterreichische Nationalbank} under project No.\ 5986 and by the 
EEC Programme {\em Human Capital and Mobility}, contract CHRX-CT93-0357 
(DG 12 COMA).

\newpage 

\begin{appendix} 
\section*{Appendix: Proper and Improper Correlators} 
We call the set of correlators 
$f_n ( t;\bk_1 , \ldots ,\bk_n )$ proper correlators if 
they cannot be written as a product of two functions,
i.e.
\beq 
f_n^{\lb \s \rb } ( t;\bk_1 , \ldots ,\bk_n ) \neq u_{m} (t; \bk_{r_1} , 
\ldots ,\bk_{r_m} ) v_{n-m} (t; \bk_{r_{m+1}} , \ldots \bk_{r_n} ) 
\eeq
where $m<n$ and $\lb r_1,\ldots ,r_n \rb $ denotes an arbitrary permutation of 
the set of indices $\lb 1,\ldots, n \rb$. In particular, proper correlators are
not the product of correlators of lower order. Conversely, improper correlators
can always be written as a product. 

We now prove that the exponentiated logarithm in the definition 
Eq.(\ref{logdef}) exactly produces all improper correlators which are 
necessary to describe a general density matrix $\rho[\be]$. Let us for the
moment drop the plus/minus index referring to starred and unstarred components
and consider all contributions to a given order $n$. (Here order is meant
to be the number of $\be$ and $\be^*$.) 
The possible types of improper graphs may be described by a partition of $n$
in the following way. Be $\lb l_1,\ldots,l_n\rb$ a set of natural numbers 
$l_\n \in \N_0$ such that $\sum_{\n=1}^{n} l_\n \n = n $, 
then this set describes an ensemble of proper correlators which contains 
$l_\n$ times -- $l_\n$ may also be zero --  a proper correlator of order $\n$,
 and which altogether represents one improper correlator of total order $n$. 
We determine the 
combinatorial factor of that improper correlator. It has $n$ momenta which 
can be attached in $n!$ permutations to the $\be$'s. However, in this way we 
over-counted  indistinguishable contractions among the momenta of the proper 
subgraphs. For each subgraph, we have to divide by the number $\n!$ of permutations 
within a subgraph. If $\lb l \rb$ contains a subgraph of order $\n$ more than once,
$l_\n > 1$, permutations of those are over-counted by a factor $l_\n !$.
Finally, we sum over all possible partitions $\lb l \rb$, which gives 
the  $n$-th order summand of $\rho [\be] $
\beq 
\rho_n [\be] = 
\sum_{\lb \s \rb} 
\sum_{ \lb l \rb} n! \prod_{\n=1}^n 
\int d \mu^+ (k_\n ) 
\frac{1}{(\n!)^{l_\n} l_\n ! }  
\be^{\s_\n} (\bk_\n ) \left( f_\n^{ \lb \s \rb} (t; \bk_i ) \right)^{l_\n} .
\eeq 
One can convince himself that restoring the $\pm$ index contained in the sum over 
$\lb \s \rb $ does not change the symmetry factors.
The complete $\be$ transform is given by the sum 
\beq 
\rho [\be] = \sum_{n=0}^\infty \frac{1}{n!}  \rho_n [\be] ,
\eeq  
where $\rho_0 [\be]=1$ encodes unit normalisation of the density operator. 
The factor $1/n!$ has to be inserted to be consistent with the definition 
(\ref{logdef}) which also involves this factor for each correlator.

In order to show that the exponentiated expression (\ref{logdef}) equals that 
expression, we write the exponentiated sum in the r.h.s\ 
as a product of single exponentials. Then, the $i$-th summand in the expanded 
exponentiated $n$-th term in the r.h.s\ of (\ref{logdef}) exactly corresponds 
to the term with $l_n = i $. The correct factor $ (n!)^{i} i ! $ in the 
denominator comes from the
$i$-th power in the Taylor-expansion of the exponential function.      
The sum over $\lb l_1 ,\ldots , l_n \rb$ just amounts to collect all terms in 
the product of the expanded exponentials which belong to order $n$. 
This completes the proof.  

\end{appendix}

\end{document}